# A Mathematical Model of Mechanotransduction

Bradley J. Roth

Department of Physics, Oakland University, Rochester, Michigan

**Abstract**: This article reviews the mechanical bidomain model, a mathematical description how the extracellular matrix and intracellular cytoskeleton are coupled by integrin proteins. The fundamental hypothesis is that differences between intracellular and extracellular displacements drive mechanotransduction. A one-dimensional example illustrates the model, which is then extended to two dimensions. In several cases the equations are solved analytically, illustrating how displacements divide into two parts: monodomain displacements are identical in both spaces and therefore do not contribute to mechanotransduction, whereas bidomain displacements cause mechanotransduction. A new length constant depends on the intracellular and extracellular shear moduli and the integrin spring constant, and bidomain effects often occur within a few length constants of the tissue edge. Numerical methods for solving the model equations are being developed. Precursors to the model and potential applications are discussed. The bidomain model may be applicable to cardiac remodeling, blood vessel regulation, tissue engineering, stem cell differentiation, cancer biology, and development.

Mailing address: Brad Roth, Dept. Physics, Oakland Univ., Rochester, MI 48309, USA
Email: roth@oakland.edu
URL: https://files.oakland.edu/users/roth/web/

**Introduction**

Suppose you want to create a mathematical model of mechanotransduction. What would your model look like? It might resemble the mechanical bidomain model (Roth 2013a). This model contains a key feature lacking in many descriptions of tissue biomechanics: the interaction of the intracellular and extracellular spaces through integrin proteins in the cell membrane. The bidomain model predicts where mechanotransduction occurs. It is a macroscopic model and therefore represents tissue averaged over many cells; it does not include the microscopic cellular structure. It describes a variety of phenomena, such as remodeling in the heart, growth of engineered tissue, stem cell differentiation, and development. The purpose of this article is to review the mechanical bidomain model and its applications.

In 1999, Matthias Chiquet wrote

> "Integrins… physically link the ECM [extracellular matrix] to the cytoskeleton, and hence are responsible for establishing a mechanical continuum by which forces are transmitted between the outside and the inside of cells in both directions…Because of their strategic location, integrins are good candidates for sensing changes in tensile stress at the cell surface … . There is evidence that upon mechanical stimulation via the ECM, integrins (or associated proteins) could trigger signals which lead to adaptive cellular responses."

Chiquet's insight suggests that integrins are responsible for initiating a cascade of molecular events that result in mechanotransduction. Many other researchers have proposed a similar role for integrins (Peyton et al. 2007; Baker and Zaman 2010; Jean et al. 2011; Kresh and Chopra 2011; Dabiri et al. 2012; Sun et al. 2012; Jacobs et al. 2013). Let us illustrate this idea with pictures and then translate it into mathematics. Figure 1 shows how integrins (red) connect the extracellular matrix (blue) to the cytoskeleton (green). What triggers the integrin's response? If the displacement in the extracellular space, **w**, differs from the displacement in the intracellular space, **u**, then the two ends of the integrin would be tugged by different amounts, causing the protein to deform. The fundamental hypothesis of our mathematical model is that the difference between the two displacements, **u** – **w**, causes mechanotransduction.

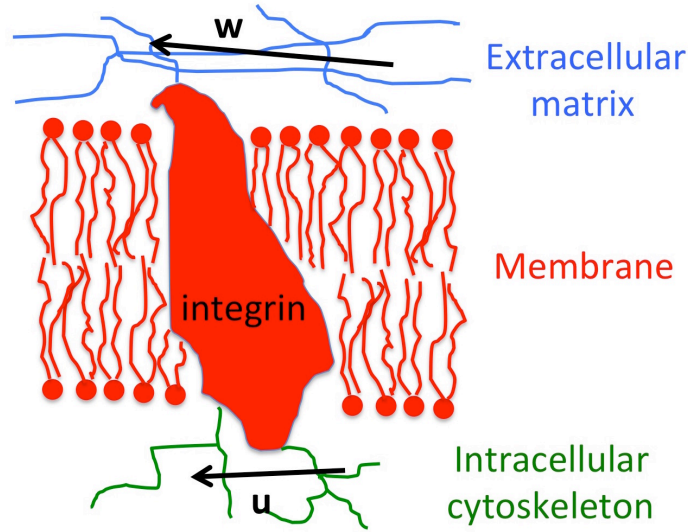

Fig. 1. The extracellular matrix interacting with the intracellular cytoskeleton through an integrin protein in the cell membrane. Differences in the intracellular and extracellular displacements, **u** and **w**, cause the integrin to deform. Our hypothesis is that such deformations cause mechanotransduction.

**Derivation of the Equations Governing the One-Dimensional Bidomain Model**

Two features of a mathematical model describing mechanotransduction are evident from Fig. 1; we need to keep track of displacements in the intracellular and extracellular spaces separately, and we must include a term representing the coupling of the two spaces by integrins. To keep things simple, consider for the moment a one-dimensional model. Assume the extracellular matrix is an elastic medium that we can represent by a line of springs (the blue springs in Fig. 2). We also represent the cytoskeleton as a line of springs (green). This representation of the intracellular space is not obvious, because tissue is made from individual cells. In order for the model to make sense, the cells need to be connected by cell-to-cell junctions, called adhesions, so when you pull on one cell the force is transferred to adjacent cells, even if the extracellular matrix is dissolved away. Finally, we represent the integrins as springs connecting the two spaces (red). The result is the ladder of springs shown in Fig. 2.

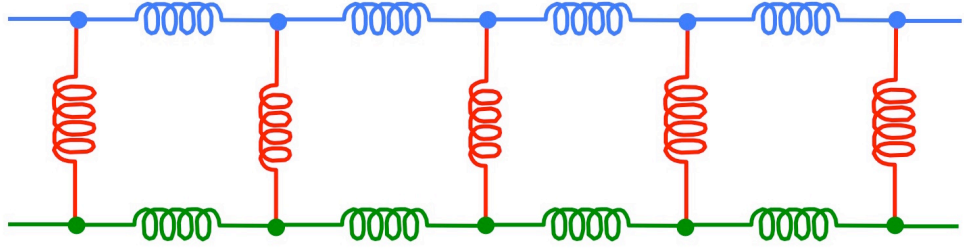

Fig. 2. The one-dimensional mechanical bidomain model. The green springs represent the intracellular space, the blue springs the extracellular space, and the red springs the integrins in the membrane.

To express this model mathematically, assume the extracellular stress, $\tau_e$, is proportional to the extracellular strain, $\varepsilon_e$, so that $\tau_e = \mu \varepsilon_e$, where μ is the extracellular mechanical modulus. The strain is simply the spatial derivative of the displacement, $\varepsilon_e = \frac{dw}{dx}$. Similarly, the intracellular stress is proportional to the intracellular strain, $\tau_i = \nu \frac{du}{dx}$, where ν in the intracellular modulus. The tissue is in mechanical equilibrium: the sum of the forces is zero. The force on any point arises from the difference between the stresses to the left and to the right of that point (in other words, the derivative of the stress), and from the force exerted by the integrins. We represent the integrins as a Hookean spring with spring constant K. This term accounts for the coupling between the two spaces, and depends on the difference between the displacements in the intracellular and extracellular spaces, u - w

$$\frac{d\tau_i}{dx} = K(u-w) \tag{1}$$

$$\frac{d\tau_e}{dx} = -K(u-w) \; . \tag{2}$$

If we put the stress-strain relationships together with the definition of the strain in terms of displacements, the equations of mechanical equilibrium become

$$\nu \frac{d^2 u}{dx^2} = K(u-w) \tag{3}$$

$$\mu \frac{d^2 w}{dx^2} = -K(u-w) \; . \tag{4}$$

Equations 3 and 4 are the one-dimensional mechanical bidomain model. The term "bidomain" means we are considering two ("bi-") spaces ("-domains"): intracellular and extracellular. The adjective "mechanical" distinguishes this model from the more familiar electrical bidomain model, which represents the electrical properties of cardiac tissue during simulations of cardiac arrhythmias and defibrillation (Henriquez 1993). We say "one-dimensional" because, as we will soon see, the model can be generalized to two and three dimensions.

The biomain equations (Eqs. 3 and 4) are a coupled pair of differential equations. To appreciate their behavior, consider what happens when you add them. The coupling terms have opposite signs (Newton's third law), so they cancel and

$$\frac{d^2}{dx^2}(\nu u + \mu w) = 0. \tag{5}$$

This is the equation of a "monodomain". That is, it is what you get for a single line of springs with a displacement given as a weighted combination of the intracellular and extracellular displacements. The integrin spring constant $K$ is not present in Eq. 5 and does not affect the monodomain behavior.

Next, divide Eq. 4 by $\mu$ and Eq. 3 by $\nu$, and then subtract them. The result is

$$\frac{d^2}{dx^2}(u-w) = \left(\frac{1}{\nu} + \frac{1}{\mu}\right) K(u-w). \tag{6}$$

We call this the "bidomain" equation for the difference in the displacements. Our fundamental hypothesis is that $u - w$ drives mechanotransduction, so this equation is crucial for understanding where mechanotransduction occurs. The equation is familiar; the solution is an exponential with length constant $\sigma$ given by

$$\sigma = \sqrt{\frac{\nu\mu}{K(\nu+\mu)}}. \tag{7}$$

The parameter $\sigma$ has units of length, and determines how rapidly the exponential falls off with distance. It is the most important parameter in the mechanical bidomain model. As the coupling spring constant $K$ gets larger, the distance $\sigma$ becomes smaller.

Equation 6 may look familiar to those who have studied bioelecticity; it is the one-dimensional cable model describing the current and voltage along a nerve axon. The steady-state cable equation is

$$\frac{d^2 V_m}{dx^2} = \frac{V_m}{\lambda^2}, \tag{8}$$

where $V_m$ is the transmembrane potential—the difference between the intracellular and extracellular potentials—and λ is the electrical length constant, which depends on the resistances of the intracellular and extracellular spaces and the membrane conductance. Many similarities exist between the electrical and mechanical models: the electrical potentials are analogous to the mechanical displacements, the electrical conductivities are analogous to the mechanical moduli, the electrical current densities are analogous to the mechanical stresses, and the electrical length constant λ is analogous to the mechanical length constant σ. In the electrical model, the opening and closing of ion channels in the membrane depends on the transmembrane potential. In the mechanical model, our hypothesis is that the activation of integrin proteins in the membrane depends on the difference between displacements; the "transmembrane displacement."

**Extension of the Bidomain Model to Two Dimensions**

We can extend the mechanical bidomain model to two or three dimensions. For instance, a two-dimensional version is shown in Fig. 3. The extracellular space (blue) and intracellular space (green) are represented by two-dimensional grids of springs coupled by integrins (red).

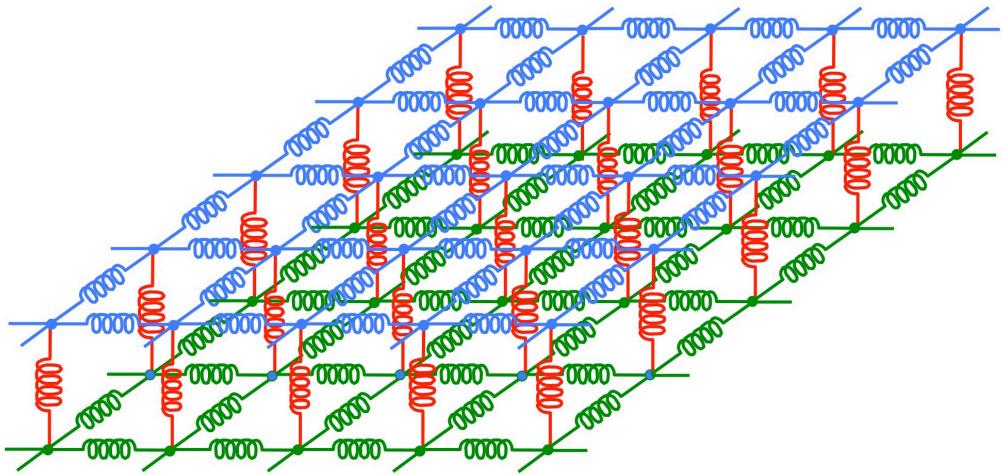

Fig. 3. The two-dimensional mechanical bidomain model. The green springs represent the intracellular cytoskeleton, the blue springs the extracellular matrix, and the red springs the integrins.

The stress-strain relationships in each space are more complicated than in one dimension because the elastic properties of a material are described by two parameters: the shear modulus and the bulk modulus (Fung 1981). Tissue, which is mostly water, is nearly incompressible. We can therefore use a hydrostatic pressure (Chadwick 1982; Ohayon and Chadwick 1988), which is the product of a tiny volume change and an enormous bulk modulus. The intracellular stress is represented by a the components of a two-dimensional tensor

$$\tau_{ixx} = -p + 2\nu\varepsilon_{ixx} \qquad \tau_{iyy} = -p + 2\nu\varepsilon_{iyy} \qquad \tau_{ixy} = 2\nu\varepsilon_{ixy}, \qquad (9)$$

where $p$ is the intracellular pressure and $\nu$ is the intracellular shear modulus. The extracellular space is similarly

$$\tau_{exx} = -q + 2\mu\varepsilon_{exx} \qquad \tau_{eyy} = -q + 2\mu\varepsilon_{eyy} \qquad \tau_{exy} = 2\mu\varepsilon_{exy}, \qquad (10)$$

where $q$ is the extracellular pressure and $\mu$ is the extracellular shear modulus. Using these stress-strain relationships, the mechanical bidomain equations become

$$-\frac{\partial p}{\partial x} + \nu\left(\frac{\partial^2 u_x}{\partial x^2} + \frac{\partial^2 u_x}{\partial y^2}\right) = K(u_x - w_x) \qquad (11)$$

$$-\frac{\partial p}{\partial y} + \nu\left(\frac{\partial^2 u_y}{\partial x^2} + \frac{\partial^2 u_y}{\partial y^2}\right) = K(u_y - w_y) \qquad (12)$$

$$-\frac{\partial q}{\partial x} + \mu\left(\frac{\partial^2 w_x}{\partial x^2} + \frac{\partial^2 w_x}{\partial y^2}\right) = -K(u_x - w_x) \qquad (13)$$

$$-\frac{\partial q}{\partial y} + \mu\left(\frac{\partial^2 w_y}{\partial x^2} + \frac{\partial^2 w_y}{\partial y^2}\right) = -K(u_y - w_y). \qquad (14)$$

The first and second equations govern the intracellular space, and the third and fourth govern the extracellular space. The first and third equations govern forces in the $x$ direction, and the second and fourth equations govern forces in the $y$ direction. Saying the tissue is incompressible is equivalent to requiring **u** and **w** have zero divergence

$$\frac{\partial u_x}{\partial x} + \frac{\partial u_y}{\partial y} = 0 \qquad \frac{\partial w_x}{\partial x} + \frac{\partial w_y}{\partial y} = 0. \qquad (15)$$

In previous analyses of the two-dimensional bidomain model, we have used stream functions to ensure incompressibility (Roth 2013a). We will not do that here, but in some cases it simplifies the analysis.

**Analytical Predictions of the Model**

The mechanical bidomain model was first derived by Puwal and Roth (2010) to describe magnetic forces on cardiac tissue. A bidomain model was necessary because the magnetic force depends on the product of the current and the magnetic field, and the intracellular and extracellular currents associated with a propagating cardiac action potential are often equal in magnitude and opposite in direction.

Therefore, in a uniform magnetic field the intracellular and extracellular forces cancel, so there is no net force on the tissue. Nevertheless, the intracellular space is pushed in one direction and the extracellular space in the other, resulting in opposite displacements. It turns out that nobody cares about magnetic forces in cardiology, but analysis of them led to the mechanical bidomain model, whose impact on the field of mechanotransduction may be far-reaching.

One prediction of the bidomain model arises from analysis of a slab of tissue being sheared (Roth, 2015). Assume that the upper surface of the slab is pulled to the right and the bottom surface to the left (Fig. 4). The displacement of the tissue is divided into two parts. The monodomain part (a weighted sum of the intracellular and extracellular displacements) varies linearly across the thickness of the slab. As a result, there is a uniform shear strain. The bidomain part (the difference between the intracellular and extracellular displacements) falls off exponentially from the upper and lower surfaces. If the length constant σ is much less than the slab thickness, then the bidomain term is negligible everywhere except near the two surfaces. In general, the bidomain part of the displacement is much smaller than the monodomain part. In Fig. 4 (and in other figures) we exaggerate the length of the bidomain arrows so they are easier to see.

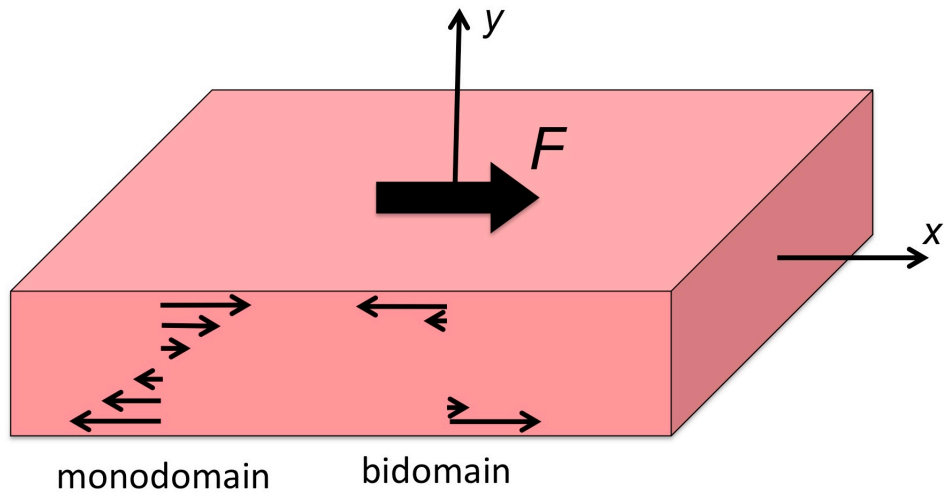

Fig. 4. The monodomain and bidomain displacements in a slab of sheared tissue; the top is pulled to the right, and the bottom to the left. Adapted from Roth (2015).

When muscle is represented by the bidomain model, we must add an additional term to the intracellular stress to account for the active tension $T$ developed by filaments of actin and myosin (Chadwick 1982; Ohayon and Chadwick 1988),

$$\tau_{ixx} = -p + 2\nu\varepsilon_{ixx} + T, \tag{16}$$

where we assume the muscle fibers are straight and aligned parallel to the *x* axis. Roth (2013b) modeled a circular sheet of cardiac tissue when the tension is uniform (Fig. 5). The sheet contracts along the fibers and incompressibility causes it to expand perpendicular to the fibers. In this case the analytical analysis is messy because the equations of elasticity are complicated in polar coordinates and the solution involves modified Bessel functions. Nevertheless, the displacement again consists of two parts: a monodomain term and a bidomain term. The monodomain strain is widely dispersed throughout the tissue. The bidomain term, however, is restricted to a thin layer near the tissue edge with a width determined by the length constant σ. Mechanotransduction occurs in this thin boundary layer.

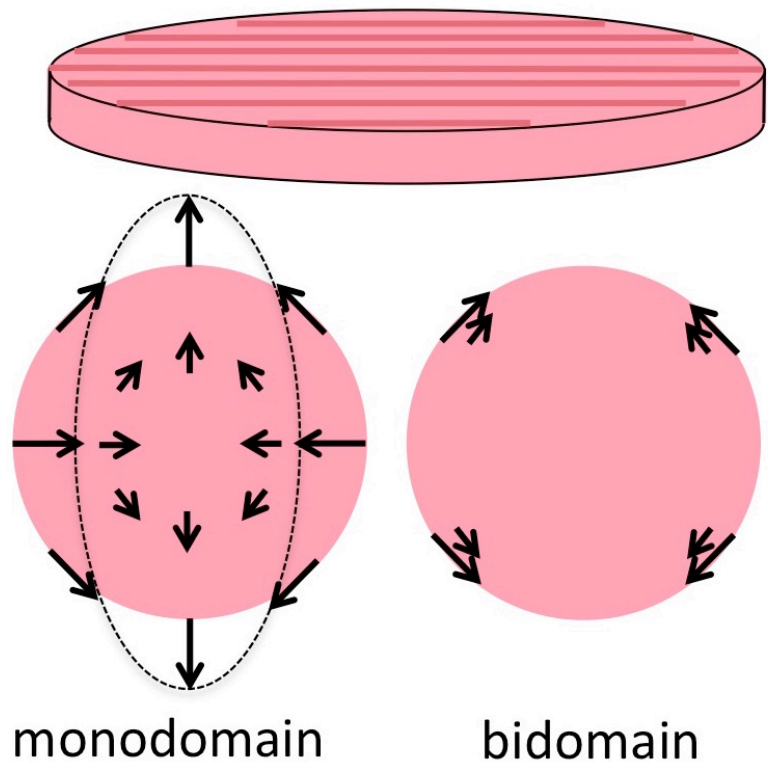

Fig. 5. Contraction of a circular sheet of cardiac tissue. The red lines indicate the fiber direction, which is horizontal in the monodomain and bidomain panels. The dotted oval in the monodomain picture shows how the sheet deforms when the fibers contract. Adapted from Roth (2013b) and from Sharma and Roth (2014).

The pressures do not vanish for the case shown in Fig. 5 (they did vanish in the example of Fig. 4), and Fig. 6 shows the intracellular and extracellular pressure distributions. The intracellular pressure is always positive, and is larger along the fiber direction than perpendicular to it. The extracellular pressure is negative along the fibers and positive perpendicular to them. The sum of the pressures is the same everywhere. The interpretation of the pressure is complicated. First, the pressure is a macroscopic quantity, and it may not be the same as the microscopic pressure

(Roth 2013a). Second, a difference between the intra- and extracellular pressures could drive water across the cell membrane. The calculation in Figs. 5 and 6 assumes the displacement occurs quickly enough that water does not have time to redistribute between spaces.

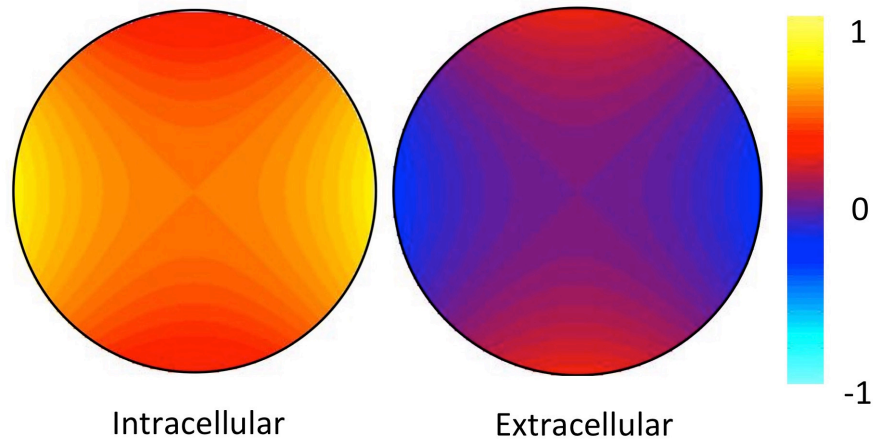

Fig. 6. The intracellular and extracellular pressures, $p$ and $q$, resulting from the contraction of muscle fibers in a circular sheet of tissue. The tissue geometry and displacements are shown in Fig. 5. The fibers are horizontal. The pressures have been normalized by their maximum value. Adapted from Roth (2013b).

   The reason pressures arise in this calculation is because we assume both spaces are incompressible. To explore this assumption further, Sharma and Roth (2014) extended the model to include both a shear modulus and a bulk modulus in each space. Because the bulk modulus allows for changes in volume, the tissue in their study was compressible. They examined several cases, including a reanalysis of the circular sheet of cardiac tissue shown in Figs. 5 and 6. They found that making the tissue compressible did not change the displacements significantly, but did change the pressures. Moreover, it introduced a second length constant into the model, similar to the first length constant (Eq. 7) except it depends on the intra- and extracellular bulk moduli rather than the shear moduli. The pressure distribution was uniform throughout most of the tissue except near the edge, where it changed over a few of these new length constants. Sharma and Roth estimate that the length constant containing the bulk moduli should be about 300 times larger than the length constant containing the shear moduli. If the bulk-modulus length constant is large compared to the dimensions of the tissue slab, then the results of the compressible model and the incompressible model are almost the same.

**Insight into the Behavior of the Mechanical Bidomain Model**

One common prediction in both Figs. 4 and 5 is the existence of a boundary layer of bidomain displacement near the tissue edge. The existence of this layer is obvious mathematically from the structure of Eq. 6, but why does it appear physically? When a force is applied to tissue, it generates a stress equal to the force divided by the tissue cross-sectional area. In the bidomain model, this stress is distributed between the intra- and extracellular spaces. For example, if the extracellular space is flexible but the intracellular space is stiff ($\mu \ll \nu$), then the stress in the extracellular space will be much smaller than the stress in the intracellular space if the displacements in the two spaces are the same. Near the tissue edge, however, the relative distribution of stresses is determined by the boundary conditions. For instance, in Fig. 4 the force $F$ is applied to the extracellular space, while the intracellular space is stress-free. The stress then redistributes between the intracellular and extracellular spaces according to the relative sizes of the shear moduli $\nu$ and $\mu$. Deep in the tissue where this redistribution is complete, the displacements and strains are the same in the two spaces although the stresses are different. The redistribution of stresses takes place over a distance of a few length constants. This analysis is similar to the "tug-of-war" mechanism describing traction forces (Trepat and Fredberg 2011). It highlights three critical features of the model: the importance of the relative size of the intracellular and extracellular shear moduli, the role of the length constant in redistributing stress between the two spaces, and the impact of the boundary conditions on the model predictions.

One well-known feature of the *electrical* bidomain model is the importance of unequal anisotropy ratios (Roth 2006). In cardiac tissue the intracellular and extracellular conductivities are similar in the direction parallel to the myocardial fibers, but the intracellular conductivity is much smaller than the extracellular conductivity perpendicular to the fibers (Roth 1997). An analogous effect may play a role in the *mechanical* bidomain model. Mechanical moduli can be anisotropic, and the anisotropy may be different in the two spaces. Earlier we pointed out the importance of the relative values of the intracellular and extracellular shear moduli in the redistribution of stresses in the tissue. If the mechanical moduli have unequal anisotropy ratios, this may lead to interesting and nonintuitive effects as stresses redistribute between the intra- and extracellular spaces when the fibers change direction. Curving and rotating fiber geometries, often encountered in the heart, may cause mechanotransduction hot spots. Cardiac monolayers can be grown with a user-specified fiber geometry (Bursac et al. 2007; Badie and Bursac 2009) that could provide a sensitive test of model predictions.

One important experiment in biomechanics is indentation (Hayes et al. 1972; Mak et al. 1987), where a probe pushes down at one point on the surface of the tissue. The probe is in direct contact with the extracellular space, and as the stresses redistribute into the intracellular space over a few length constants around the probe we expect a region where there are large differences between **u** and **w**, resulting in mechanotransduction. However, unequal anisotropy ratios may imply that this redistribution has a complicated and surprising spatial pattern, much like the unexpected spatial distribution of transmembrane potential around a stimulating electrode predicted by the electrical bidomain model (Sepulveda et al. 1989).

Another relatively unexplored aspect of the mechanical bidomain model is the relationship between the macroscopic model and the microscopic tissue structure. Again an analogy exists between the mechanical and the electrical models. One hypothesis for how electric shocks affect the heart is the sawtooth model, which is a microscopic model that separates the electrical resistance of the cytoplasm from the resistance of the gap junctions coupling cells (Plonsey and Barr 1986; Krassowska et al 1987). Gap junctions in the electrical bidomain model are analogous to adhesions (mechanical intercellular junctions) in the mechanical bidomain model. The forces acting on the adhesions may lead to mechanotransduction (McCain et al. 2012). Such considerations go beyond the macroscopic bidomain model and explore the macroscopic/microscopic relationship. Mertz et al. (2013) have found that biomechanical behavior is sensitive to cell-cell adhesions, and this sensitivity may provide another tool with which to probe mechanotransduction.

One difference between the electrical and mechanical bidomain models is that the transmembrane potential is a scalar quantity, whereas the bidomain displacement is a vector. Moreover, in the electrical model a positive transmembrane potential (depolarization) has a different effect than a negative transmembrane potential (hyperpolarization). Our working hypothesis has been that the magnitude of **u** – **w** is the key quantity in the mechanical bidomain model, not its sign or direction. But we don't know this for sure, and perhaps its direction is important too.

**Numerical Calculations Using the Model**

Few biomechanics problems can be solved analytically; most require numerical analysis. Punal and Roth (2012) analyzed the mechanical bidomain model using perturbation theory (Johnson 2005). If there is distance characterizing the tissue, such as the thickness of the tissue slab (Fig. 4) or radius of the tissue sheet (Fig. 5), we can form a dimensionless parameter by dividing the bidomain length constant $\sigma$ by the characteristic distance. In most cases, we expect $\sigma$ will be much smaller than the other length scale, so this dimensionless parameter will be small. Punal and Roth expanded their expressions in powers of this small parameter and then collected terms with like powers. Their zeroth order equations governed the lowest-order monodomain contribution and the first order equations governed the first nonzero bidomain contribution. Recall that only the bidomain term contributes to mechanotransduction (the monodomain term results in identical displacements in the intracellular and extracellular spaces, so it does not contribute to their difference). Thus, perturbation theory could provide a two-step process for numerical biomechanics: first solve the monodomain equation just like everyone else in the field of biomechanics does, and then use this solution and the first order equation to calculate the bidomain contribution. This technique would be valuable, because it would tie the bidomain model to monodomain biomechanical models that preceded it (Guccione and McCulloch 1991; Vetter and McCulloch 1998; McCulloch 2006; Humphrey 2010). Monodomain models often contain important features not yet included in the bidomain model, such as large strains, nonlinear stress-strain relationships, and complicated tissue and fiber geometries. Perturbation methods may provide a way to predict bidomain displacements from previous sophisticated, nonlinear monodomain simulations.

Another way to analyze the mechanical bidomain model is to solve the equations numerically on a computer using the finite difference method. Gandhi and Roth (2016) developed such a technique to study remodeling of tissue around an ischemic region in the heart (Fig. 7). The central circular area is ischemic and cannot develop a tension. The surrounding tissue is healthy and has a uniform tension $T$ acting along the myocardial fibers (horizontal). When the healthy tissue contracts, it stretches the ischemic region. Because of incompressibility, the ischemic area is flattened in the direction perpendicular to the fibers. A complex distribution of monodomain strain extends throughout the ischemic area and the surrounding healthy tissue. The bidomain displacement, however, is confined to the ischemic region's border zone. The model predicts that remodeling of cardiac tissue—a type of mechanotransduction—should occur primarily in the border zone, consistent with observations Rodriguez et al. (2005).

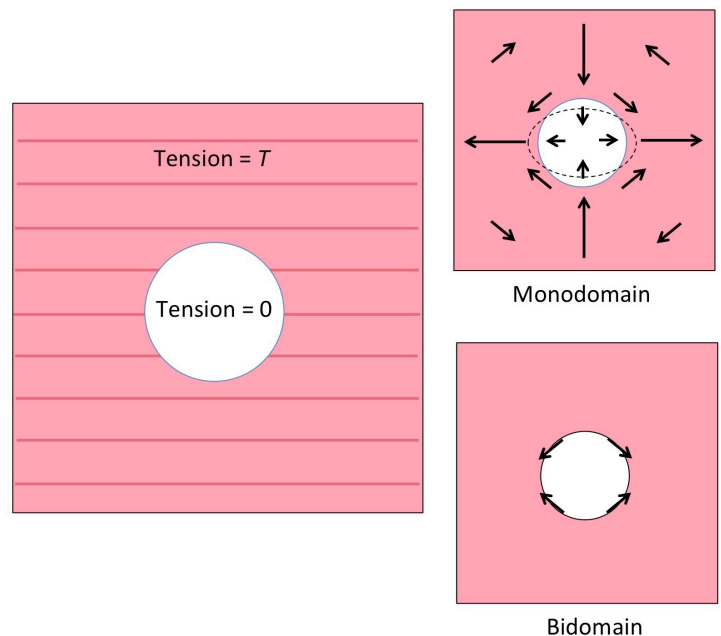

Fig. 7. A sheet of cardiac tissue with a central ischemic region that cannot develop an active tension. Adapted from Gandhi and Roth (2016).

Sharma et al. (2015) have performed the first finite element calculation using the bidomain model. Such calculations are important, because the finite element model is necessary in order to apply biomechanical models to tissues with realistic and complicated geometries (Guccione and McCulloch 1991; Nash and Hunter 2000). Our long-term goal is to create a bidomain model of the whole heart.

The mechanical bidomain model is an example of a multiscale model (De et al. 2015): the length scales of interest extend from molecules to cells to tissue and finally to the organ, spanning spatial scales of many orders of magnitude (Fig. 8).

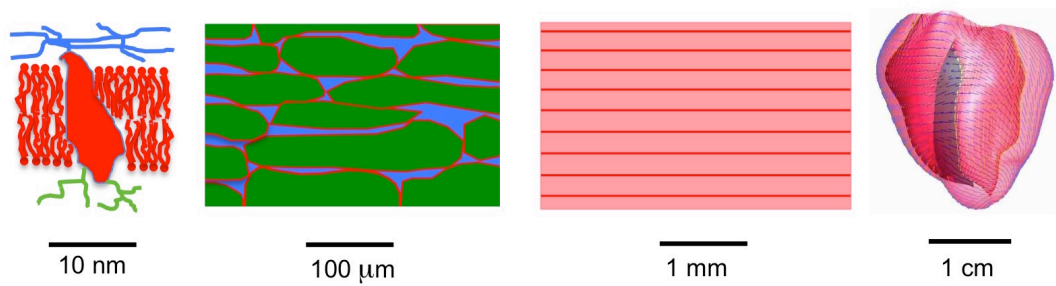

Fig. 8. The multiscale nature of the mechanical bidomain model. The length scales range from the molecular (an integrin in the cell membrane) to the cellular (cardiac cells embedded in extracellular matrix and surrounded by a cell membrane) to the tissue (the macroscopic mechanical bidomain model of cardiac tissue with red lines showing the fiber direction) to the organ (a model of a rabbit heart). The rightmost panel is modified from Vetter and McCulloch (1998).

**Precursors of the Bidomain Model**

The mechanical bidomain model grew out of several earlier studies. First and foremost is the electrical bidomain model (Henriquez 1993), which is now the state-of-the-art model for simulating pacing and defibrillation of the heart (Trayanova and Plank 2009). The mechanical and electrical bidomain models have many similarities, including their mathematical structure. One rationale for developing the mechanical bidomain model is that it may become as important for studies of mechanotransduction as the electrical bidomain model is for studies of defibrillation.

Mechanical models similar to the bidomain model have been proposed previously, and are generally called "biphasic" models. Perhaps the best known is Mow's biphasic model of cartilage (Mow et al 1984; Mak et al 1987; Ateshian et al 1997). The solid and fluid phases in cartilage are analogous to the intra- and extracellular spaces in cardiac tissue, and the frictional coupling of cartilage's two phases is governed by a mathematical term that is similar in form to the elastic coupling of the intra- and extracellular spaces by integrins in the bidomain model.

The mechanical bidomain model has many similarities with the models derived by Edwards and Schwarz (2011) and Banerjee and Marchetti (2012) to describe growing cell colonies (Mertz et al. 2012; 2013). Edwards and Schwarz's spring constant $k$ is analogous to our constant $K$, and their localization length $l$ is similar to our length constant $\sigma$. However, they considered the intracellular space coupled to a microstructured surface consisting of an array of flexible elastomeric pillars, as is often used in traction force experiments (Style et al. 2014). Our model, on the other hand, interprets this coupling as occurring via integrins. Therefore our coupling term takes on a different significance than in previous models: in our model it is the signal that drives mechanotransduction.

**Potential Applications of the Mechanical Bidomain Model**

Many potential applications of this model exist. We have already mentioned remodeling of cardiac tissue in the heart. Not only can the model predict tissue changes in the border zone of an ischemic region, but also it might explain the thickening of the whole heart during hypertrophy. Puwal (2013) has begun using the bidomain model to predict how the heart responds to elevated blood pressure and other abnormalities. Kroon et al. (2009) and Bovendeerd (2012) have suggested that mechanical stimuli and fiber orientation may impact growth and remodeling in the heart. They postulated that ventricular wall stress may be the stimulus for such mechanotransduction, but our model suggests that differences in intracellular and extracellular displacements may be the driver of these events, implying that a bidomain formulation is essential for studying remodeling.

The mechanical bidomain model could be useful for predicting the opening of stretch-activated ion channels. Mechanoelectrical feedback in cardiac tissue (Kohl and Sachs 2001) may be responsible for stretch-induced arrhythmias (Hansen et al 1990) and could impact defibrillation efficacy (Trayanova et al 2004; Li et al 2008). The activation mechanism of stretch-activated ion channels is unclear; these ion channels may respond to membrane forces, or they may be controlled by stretch sensors in the intracellular space (Knoll et al 2002). The mechanical bidomain model might indicate how to distinguish between these two hypotheses.

Shear forces play an important role in the physiology of a blood vessel (Pan 2009; Lu and Kassab 2011). The vascular endothelium is regulated by shear stress caused by blood flow (Chiu and Chien 2011), in part through production of nitric oxide (Balligand et al 2009). A model of a blood vessel and the blood flowing within it will allow us to explore the impact of the mechanical bidomain model on this behavior. Such simulations would require deriving the appropriate boundary conditions to couple the bidomain tissue to blood flow.

Engineered tissue is becoming increasingly important for therapy (Zimmermann et al 2004; Naito et al 2006; Butler et al 2009). Tissue engineering in general often requires careful manipulation of mechanical forces (Guilak et al. 2014). *In vitro* tissue engineering relies on the prefabrication of replacement tissue (Bach et al

2003) grown on an extracellular matrix (Silva and Mooney 2004). The mechanical stresses that the tissue experiences during growth influence its structure and function (Powell et al 2002; Katare et al 2010). For example, Fink et al. (2000) stretched a sample of engineered heart tissue and found a greater concentration of cells embedded in the tissue at the edge. While we have not yet modeled this experiment in detail, the localization of tissue growth at the edge is suggestive of a bidomain effect.

Mechanical forces play a key role in controlling tissue growth (Sun et al. 2012) and stem cell differentiation (Yim and Sheetz 2012). In a colony of growing cells, tissue properties are often different at the periphery than in the interior (Nelson et al. 2005; Ruiz and Chen 2008; Mertz et al. 2012). For example in colonies of growing human stem cells, traction forces and differentiation occur primarily at the edge of the colony (Rosowski et al. 2015). The mechanical bidomain model also predicts that mechanotransduction occurs at the colony edge. This data may be the most compelling yet in support of the bidomain approach. Moreover, Rosowski et al.'s study provides an estimate of the size of the length constant $\sigma$. They observe edge effects that extend over about 150 microns, which is larger than the size of their individual cells but smaller than the radius of their colonies. If the mechanical bidomain model correctly predicts the behavior of stem cell colonies, it might provide insight into the complex process of human development.

Finally, evidence exists that integrins may play a role in tumor biology and cancer therapy (Baker et al. 2009; Jean et al. 2011). The growth of tumors might therefore be impacted by bidomain effects.

**Conclusion**

The mechanical bidomain model is still in the early stages of its development. Many of the applications discussed in the last section have not yet been analyzed, and the model may not prove fruitful in each case. Nevertheless, several authors have claimed that integrin coupling of the cytoskeleton and the extracellular matrix plays an important role in mechanotransduction. The mechanical bidomain model represents a first step in developing a mathematical model of this interaction. Indeed, it is almost the simplest model one could derive that includes the coupling of the two spaces by integrins. Certainly additional factors will need to be added to the model when applying it to different cases. But even in its simplest form, the model provides valuable insight into mechanotransduction. Finally, if the model predictions prove to be inconsistent with experiments, the process of developing the model will still be useful because it will force researchers to analyze why the model is incorrect. By clarifying what aspect of the model must be modified or eliminated before it accurately predicts experimental data, we gain insight into the mechanisms of mechanotransduction.